\documentclass[aps,pra,twocolumn,amssymb,amsfonts,floatfix,showpacs,]{revtex4}
\usepackage{bm}
\usepackage{dcolumn}
\usepackage{bm}
\usepackage{mathbbol}
\usepackage{graphicx}
\usepackage{epstopdf} 
\usepackage{amsmath,amsthm}
\usepackage{times}
\usepackage{color}
\usepackage{mathtools}

\begin{document}
\title{Dynamics at the Many-Body Localization Transition}

\author{E. J. Torres-Herrera$^{1,2}$ and Lea F. Santos$^{1,3}$} 
\affiliation{$^1$ Department of Physics, Yeshiva University, New York, New York 10016, USA \\ 
$^2$ Instituto de F{\'i}sica, Universidad Aut\'onoma de Puebla, Apt. Postal J-48, Puebla, Puebla, 72570, Mexico \\
$^3$ ITAMP, Harvard-Smithsonian Center for Astrophysics, Cambridge, Massachusetts 02138, USA}

\date{\today}

\begin{abstract}
The isolated one-dimensional Heisenberg model with static random magnetic fields has become paradigmatic for the analysis of many-body localization. Here, we study the dynamics of this system initially prepared in a highly-excited nonstationary state. Our focus is on the probability for finding the initial state later in time, the so-called survival probability.  Two distinct behaviors are identified before equilibration. At short times, the decay is very fast and equivalent to that of clean systems. It subsequently slows down and develops a powerlaw behavior with an exponent that coincides with the multifractal dimension of the eigenstates. 
\end{abstract}


\maketitle


\section{Introduction} 
The metal-insulator transition has been at the forefront of physics research since Anderson's seminal paper~\cite{Anderson1958}. As a result of quantum interference, the wavefunctions of a disordered noninteracting system can become exponentially localized in configuration space. The phenomenon has been experimentally observed in different setups, more recently with Bose-Einstein condensates~\cite{Billy2008,Roati2008}. A proposal for an experiment with ultracold atoms in a two-dimensional geometry also exists~\cite{MorongARXIV}. Lattice models, such as the Anderson tight-binding and the powerlaw random banded matrix (PRBM) models~\cite{Evers2008,notePRBM}, have been extensively employed in the analysis of the Anderson metal-insulator transition. At criticality, it was found that the eigenstates exhibit multifractal features~\cite{Evers2008}.

The inverse participation ratios, $\text{IPR}_q^{\alpha} = \sum_n |C^{\alpha}_n|^{2q}$, contain information about the structure of the eigenstates $|\psi^{\alpha} \rangle = \sum_n  C^{\alpha}_n |\phi_n \rangle$ written in the basis vectors $|\phi_n \rangle$ of the configuration space. In particular, $\text{IPR}_2^{\alpha}$ measures the level of delocalization of the eigenstates in the chosen basis~\cite{notePR}. At the Anderson transition, the probability amplitudes $C^{\alpha}_n$ display large fluctuations and $\text{IPR}_q^{\alpha}$
shows anomalous multifractal scaling with respect to the system size~\cite{Wegner1980,Soukoulis1984,Castellani1986,Evers2000,Evers2001,Evers2008},
\begin{equation}
\langle \text{IPR}_q^{\alpha}  \rangle  \sim {\cal N}^{-(q-1) D_q},
\label{eq:Generalized}
\end{equation}
where $\langle . \rangle$ denotes the average over an ensemble of realizations and eigenstates,  ${\cal N}$ is the dimension of the Hamiltonian matrix~\cite{noteAnderson}, and $D_q$ represents the generalized dimension. Multifractality is reflected by the nonlinear dependence of the generalized dimension on $q$.  In contrast, $D_q=d$ in the metallic phase, where $d$ is the system dimension, and $D_q=0$ in the insulating phase. Experimentally, multifractality has been observed in disordered conductors~\cite{Richardella2010} and in systems with cold atoms~\cite{Lemarie2010,Sagi2012}. Recently, new studies have led to the conclusion that multifractal correlations are not exclusive to the critical point of the Anderson-transition. In disordered systems, they are present away from criticality~\cite{Falko1995} and even in extended states~\cite{DeLuca2014}. They are also found in the ground states of clean systems~\cite{Atas2012}.

Studies of the dynamics of noninteracting systems at the metal-insulator transition have shown that the Loschmidt echo~\cite{Ng2006},  the survival probability~\cite{Ketzmerick1992,Huckestein1994,Huckestein1999}, and the spreading of wavepackets~\cite{Ketzmerick1997,Huckestein1999} at the mobility edge exhibit a powerlaw behavior, where the exponent coincides with the generalized dimension for $q=2$. The generalized dimension of the eigenstates, $D_2$, is extracted from Eq.~(\ref{eq:Generalized}) by performing a scaling analysis of $\langle \text{IPR}_2^{\alpha}\rangle$. However, in studies of dynamics the main interest is on the generalized dimension associated with the initial state $|\phi_{n_0} \rangle$ and denoted by $\widetilde{D}_2$. The latter is obtained from a scaling analysis of the level of delocalization of the initial state with respect to the energy eigenbasis, that is the analysis of
\begin{equation}
\langle \text{IPR}_2^{n_0}  \rangle  \sim {\cal N}^{-\widetilde{D}_2},
\label{eq:D2forInitial}
\end{equation}
where $|\phi_{n_0} \rangle = \sum_{\alpha} C^{\alpha}_{n_0} |\psi^{\alpha} \rangle$ and $\text{IPR}_2^{n_0} =  \sum_{\alpha} |C^{\alpha}_{n_0} |^4$.
When investigating localization in real space, the initial state usually corresponds to a basis vector of the configuration space. We also note that the two generalized dimensions above have been shown~ \cite{Huckestein1997} to be related through the expression $D_2= d \widetilde{D}_2$.

A natural question following this brief summary of the Anderson localization is what happens to the above findings when interaction is included. It had been conjectured already in~\cite{Anderson1958,Fleishman1980} and then confirmed with perturbative arguments~\cite{Gornyi2005,Basko2006} and rigorously~\cite{ImbrieARXIV} that localization may persist. 
Studies about many-body localization (MBL) have recently boomed~\cite{SantosALL,Pal2010,DeLuca2013,Kjall2014,Luitz2014,Oganesyan2007,Znidaric2008,Bardarson2012,Vosk2013,Serbyn2014,Lev2014,LevARXIV,AgarwalARXIV,VasseurARXIV,Zangara2013PRB,Aleiner2010,Huse2013,Serbyn2013,Chandran2014,Huse2014,Ros2015,Serbyn2014PRL,TangARXIV}. The interest in the subject is in part motivated by the access to new experimental tools, such as cold atoms in optical lattices~\cite{SchreiberARXIV}, that can be used to corroborate theoretical predictions. Among the latter, we find works about the location of the critical point in disordered spin-1/2 chains~\cite{Pal2010,DeLuca2013,Kjall2014,Luitz2014}, analysis of the relation between the distribution of the wavefunction coefficients and the onset of localization~\cite{DeLuca2013}, various efforts to identify the quasi-local integrals of motion in the MBL phase~\cite{Serbyn2013,Chandran2014,Huse2014,Ros2015},  and descriptions of the evolution of the entanglement entropy~\cite{Znidaric2008,Bardarson2012}, few-body observables~\cite{Serbyn2014,Lev2014,LevARXIV,AgarwalARXIV,VasseurARXIV}, and the Loschmidt echo~\cite{Zangara2013PRB}.

Our goal in this work is to characterize the evolution of isolated disordered systems with interaction from very short to very long times. Since MBL is a dynamical transition, identifying general features of the dynamics of interacting systems is essential for the further developments of the field. Motivated by the results for noninteracting systems, our focus is on the decay of the survival probability and its relationship with the onset of multifractal states.

We consider a one-dimensional (1D) disordered spin-1/2 system and analyze the evolution at different time scales of both the survival probability and the time-averaged survival probability. At short times the decay is very fast and similar to that of clean systems. Afterwards, the decay slows down and shows an anomalous powerlaw behavior. The exponent of this algebraic decay coincides with $\widetilde{D}_2$. At very long-times, the decay eventually saturates to $\langle \text{IPR}_2^{n_0}  \rangle$.


\section{ Model and Basis} 
We investigate the 1D isotropic Heisenberg spin-1/2 system with two-body nearest-neighbor interaction, $L$ sites, and periodic boundary conditions. The Hamiltonian is 
\begin{equation}
\widehat{H}= \sum_{k=1}^L \left[ h_k  \widehat{S}_k^z  + J \left(
\widehat{S}_k^x \widehat{S}_{k+1}^x + \widehat{S}_k^y \widehat{S}_{k+1}^y +\widehat{S}_k^z \widehat{S}_{k+1}^z \right) \right] .
\label{ham}
\end{equation}
Above $\hbar =1$, $\widehat{S}^{x,y,z}_k $ are spin operators, and $J=1$ sets the energy scale. Random static magnetic fields act on each site $k$, the amplitudes $h_k$ being random numbers from a uniform distribution $[-h,h]$. The total spin in the $z$-direction, $\widehat{{\cal{S}}}^z=\sum_k\widehat{S}_k^z$, is conserved. We work with the largest subspace, ${\cal{S}}^z=0$, of dimension ${\cal N}=L!/(L/2)!^2$. Localization in this symmetry sector guarantees localization in smaller sectors. 

The dependence on $h$ of the level statistics and of the level of delocalization of the eigenstates of $\widehat{H}$ (\ref{ham}) has been studied for at least a decade~\cite{Avishai2002,SantosALL,Pal2010,DeLuca2013,Luitz2014}. When $h=0$, the system is analytically solvable with the Bethe ansatz. If all the trivial symmetries of the Hamiltonian are taken into account, one verifies that the level spacing distribution of neighboring levels is Poisson. In addition to the total spin in the $z$-direction,  the other symmetries of the isotropic model at  ${\cal{S}}^z=0$ are: translational invariance, parity, spin reversal, and conservation of total spin. 

As $h$ increases from zero, the level spacing distribution eventually becomes Wigner-Dyson, indicating a transition to the chaotic regime. The value of $h$ at which level repulsion becomes evident decreases as the system size increases. In parallel, the level of delocalization of the eigenstates in real space increases substantially. The presence of disorder breaks the additional symmetries mentioned above and if the disorder is weak the states can spread out significantly. 

As $h$ further increases and becomes larger than the coupling strength, $h>1$, the spreading of the eigenstates recedes and they become more localized in real space.  The critical point for the transition to the MBL phase has been identified as $h_c \cong 3.5 \pm 1.0$ in~\cite{Pal2010} and $h_c \cong 2.7 \pm 0.3$ in~\cite{DeLuca2013}. 

In any study of the structure of the eigenstates, the choice of basis is essential. Since here the goal is to investigate localization in real space, that is the level of confinement of the spin excitations in the lattice, the natural basis is that of the configuration space, which we refer to as the site-basis and is also known as the computational basis.  The site-basis vectors $|\phi_{n} \rangle$ correspond to states where the spin on each site either points up or down along the $z$-axis, such as $|\uparrow \downarrow \uparrow \downarrow \ldots\rangle $.


\section{Survival Probability}  
To study the dynamics of the disordered chain (\ref{ham}), we take as initial state a single site-basis vector,   $|\Psi(0)\rangle = |\phi_{n_0} \rangle$. This is equivalent to a quench, where the initial Hamiltonian is the Ising part of the Hamiltonian $\sum_{k=1}^L \widehat{S}_k^z \widehat{S}_{k+1}^z $ and the final Hamiltonian is  $\widehat{H}$ (\ref{ham}).
To quantify how fast the initial state changes in time, we concentrate on the behavior of the survival probability,
\begin{equation}
F(t) =\left| \langle \Psi(0) | e^{-i \widehat{H} t} | \Psi(0) \rangle \right|^2  
= \left|\sum_{\alpha} |C^{\alpha}_{n_0} |^2 e^{-i E_{\alpha} t}  \right|^2 ,
\label{eq:fidelity}
\end{equation} 
where $E_{\alpha} $ are the eigenvalues of $\widehat{H}$ and $C^{\alpha}_{n_0}=\langle \psi^{\alpha} |\phi_{n_0} \rangle$ is the overlap of the initial state with the eigenstates $|\psi^{\alpha} \rangle $ of $\widehat{H}$. $F(t)$ measures the probability for finding the system still in $|\Psi(0)\rangle $ at time $t$. 

The distribution in energy 
\begin{equation}
\rho_{n_0}(E) = \sum_\alpha |C^{\alpha}_{n_0}|^2\delta(E-E_\alpha)
\end{equation}
 of the components $|C^{\alpha}_{n_0}|^2$ of the initial state is often referred to as local density of states (LDOS). If the envelope of this distribution is known, an analytical expression for $F(t)$ can be obtained from the Fourier transform, $F(t) \simeq \int \rho_{n_0}(E) e^{-i E t} dE$.

For strong quenches, that is when the initial and final Hamiltonians are very different, the envelope of $\rho_{n_0}(E)$ is a Gaussian with mean corresponding to the energy of the initial state, 
\begin{equation}
\varepsilon_{n_0} = \sum_{\alpha} |C^{\alpha}_{n_0} |^2 E_{\alpha} =\langle \phi_{n_0} |\widehat{H} |\phi_{n_0} \rangle 
\label{eq:mean}
\end{equation}
and width  
\begin{equation}
 \sigma_{n_0}^2 = \sum_{\alpha} |C^{\alpha}_{n_0} |^2 E_{\alpha}^2 - \varepsilon_{n_0}^2 =\sum_{n \neq n_0} |\langle \phi_n |\widehat{H} |\phi_{n_0} \rangle|^2.
 \label{eq:width}
\end{equation}
 Notice that the width depends only on the off-diagonal elements of the Hamiltonian matrix written in the site-basis and is therefore independent of the diagonal disorder. The Gaussian shape of the LDOS reflects the density of states, which for systems with two-body interaction is also Gaussian~\cite{Brody1981}.
  
In the absence of disorder, the envelope $\rho_{n_0}(E)$ is particularly well filled for initial states with energy $\varepsilon_{n_0} $ near the center of  the spectrum of $\widehat{H}$ \cite{Zangara2013,TorresALL}.  Its Gaussian shape leads to the Gaussian decay $F(t) \sim \exp(-\sigma_{n_0}^2 t^2)$. This behavior may persist until saturation or be followed by an exponential~(see Refs.~\cite{TorresALL,Izrailev2006} and references therein). 
  
\begin{figure}[htb]
\centering
\includegraphics*[width=3.in]{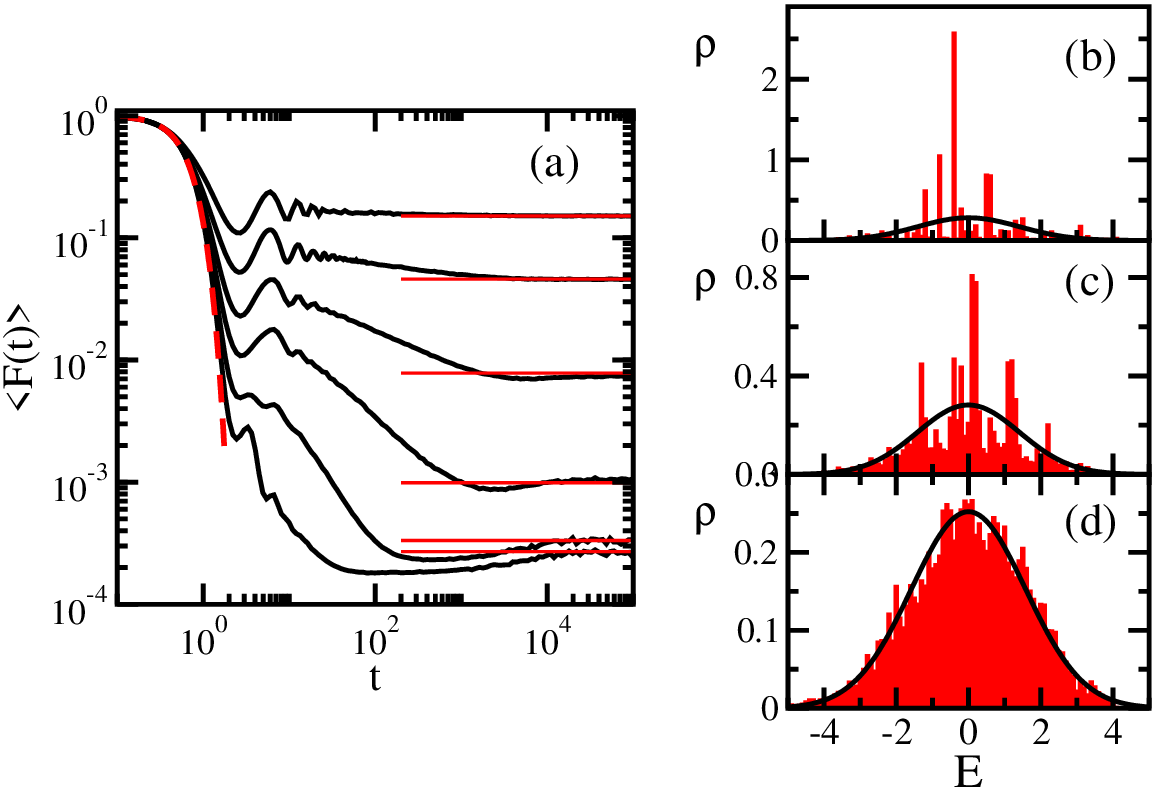}
\caption{(Color online) Survival probability averaged over $10^5$ data points for $h=0.5, 1.0,1.5,2.0,2.7,4.0$ from bottom to top (a) and LDOS for a single realization for the bottom panel $h=0.5$ (d), the middle $h=1.5$ (c), and the top $h=2.7$ (b); $L=16$. In (a): the dashed line indicates $\exp(-\langle \sigma_{n_0}^2\rangle t^2)$ for $h=0.5$ and horizontal lines correspond to the saturation point, $\langle \text{IPR}_2^{n_0} \rangle$. The envelopes (solid lines) of the distributions in panels (b), (c), and (d) are Gaussians with center $\varepsilon_{n_0}$ [Eq.~(\ref{eq:mean})] and width $\sigma_{n_0}$ [Eq.~(\ref{eq:width})].
}
\label{fig:LDOS}
\end{figure}

In Fig.~\ref{fig:LDOS} we analyze the survival probability and the LDOS in the presence of disorder.
The average of $F(t)$ over different disorder realizations and different initial states is denoted by $\langle F(t) \rangle$. For each system size and each realization, we select as initial states, only 10\% of all the ${\cal N}$ site-basis vectors. They are the ones with energy $\varepsilon_{n_0} $ closest to the middle of the spectrum of $\widehat{H}$. Since the density of states is Gaussian, the center of the spectrum contains the most delocalized states. Localization in this region assures localization in other parts of the spectrum. For each $L$, the total number of data points for the average, including initial states and realizations, adds up to $10^5$.

Figure~\ref{fig:LDOS} (a) displays $\langle F(t) \rangle$ for different values of $h$. 
The initial decay is very fast until $t\sim 2$. For  small disorder, the initial evolution is purely Gaussian. As $h$ increases, the interval of the Gaussian decay shrinks until only the quadratic part persists, $\langle F(t) \rangle \sim 1 - \langle \sigma_{n_0}^2 \rangle t^2$. This is followed by a possible exponential behavior, the time interval being too short for certainty. After the initial fast evolution, oscillations appear. The time interval of these oscillations as well as their amplitudes increase with the disorder strength. 

The oscillations eventually fade away and give place to a powerlaw decay with exponent $\leq 1$. The initial state finds new channels that give continuation to its evolution. The couplings at higher order in perturbation theory become gradually effective. 

The long-time powerlaw behavior reflects the onset of multifractal states~\cite{Ketzmerick1992,Huckestein1994,Huckestein1999}. Our results indicate that multifractal many-body states can occur even at small $h$.   As the disorder increases and the eigenstates become less extended, the powerlaw exponent naturally decreases. For $L=16$, the decay after the oscillations is hardly noticeable for $h \gtrsim 4$. 

At very long times, the decay eventually saturates. The saturation point is derived from the infinite time average of Eq.(\ref{eq:fidelity}),
\[
\langle F(t\rightarrow \infty) \rangle \sim \left\langle  \sum_{\alpha} |C^{\alpha}_{n_0} |^4 \right\rangle =  \langle  \text{IPR}_2^{n_0} \rangle. 
\]
The value of this infinite time average naturally increases with the disorder strength.

Figures~\ref{fig:LDOS} (b), (c), and (d) display representative LDOS for three values of $h$. The widths of the three distributions are equivalent, because according to Eq.~(\ref{eq:width}), $\sigma_{n_0}$ does not depend on the disorder strength. This explains the indistinguishable initial decay for all curves in Fig.~\ref{fig:LDOS} (a). 

At small $h$ [Fig.~\ref{fig:LDOS} (d)], the Gaussian envelope of the distribution is still well filled, indicating a very delocalized initial state. This is independent of the realization, provided $\varepsilon_{n_0} $ be near the center of the spectrum.  As the disorder increases, the multifractal structures of the eigenstates spread to larger scales and the coefficients $C^{\alpha}_{n_0}$ fluctuate strongly. As a result, the LDOS becomes more sparse [Figs.~\ref{fig:LDOS} (b) and (c)], justifying the oscillations and subsequent powerlaw decay in Fig.~\ref{fig:LDOS} (a). The oscillations are due to the small number of states energetically accessible to the initial state in low order of perturbation theory; a number that decreases as $h$ increases. These oscillations are not random fluctuations that can be averaged out with enough realizations, as those at very long times. They are connected with the approach to the MBL phase and the onset of quasi-integrals of motion~\cite{Serbyn2013,Chandran2014,Huse2014,Ros2015}.

\begin{figure}[htb]
\centering
\includegraphics*[width=3.in]{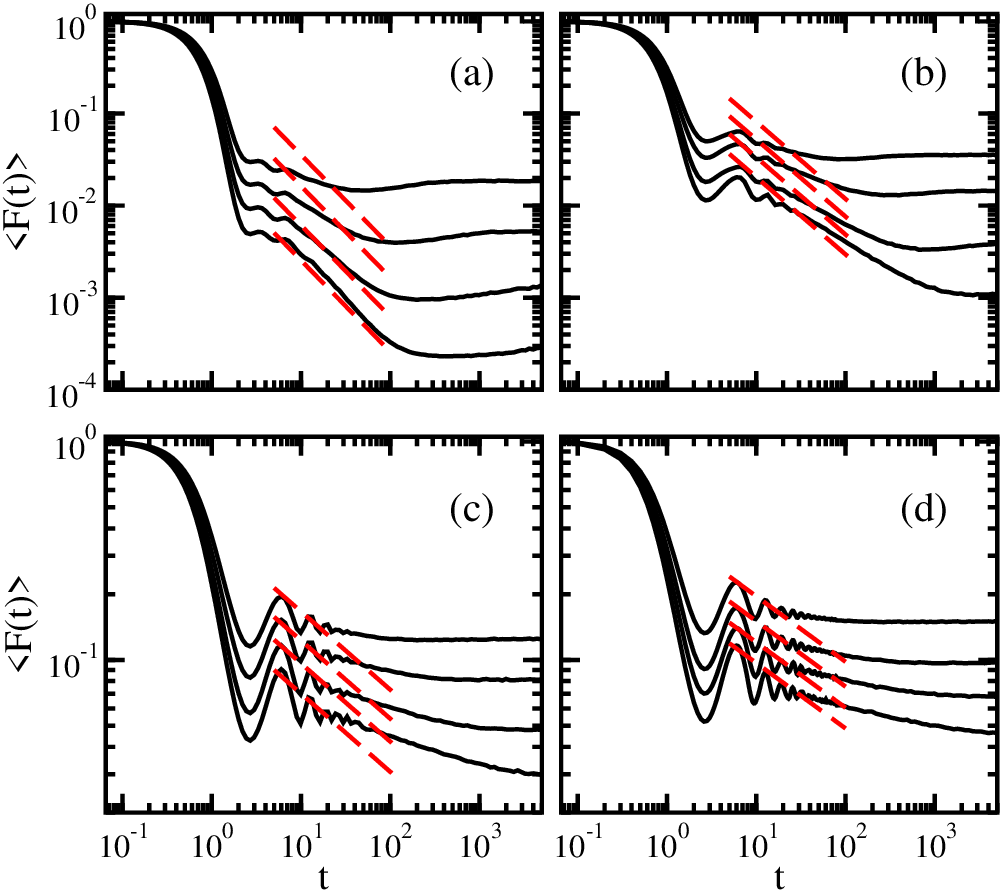}
\caption{(Color online) Survival probability averaged over $10^5$ data point for $h=1.0$ (a),  $h=1.5$ (b), $h=2.5$ (c), and $h=2.7$ (b) for $L=10, 12, 14, 16$ from top to bottom. Dashed lines give $t^{-\widetilde{D}_2}$, where $\widetilde{D}_2 = 0.99$ (a), $0.84$ (b), $0.36$ (c), and $0.30$ (d). 
}
\label{fig:SPoneCase}
\end{figure}

In Fig.~\ref{fig:SPoneCase}, we analyze the survival probability for different system sizes and four values of $h$.  The strengths of the disorder are small in Figs.~\ref{fig:SPoneCase} (a), (b), while in Figs.~\ref{fig:SPoneCase} (c), (d), they coincide, within errors, with the critical point $h_c$ obtained in Refs.~\cite{Pal2010,DeLuca2013}. The fast evolution for $t<2$ is separated from the later powerlaw decay either by a small plateau (a) or by visible oscillations (c), (d). The scope of the powerlaw behavior increases with system size and with disorder strength[compare the time where saturation takes place in (a) with the time in (c), for example]. This suggests that for very large $L$ the algebraic decay may persist for $h>h_c$.

\begin{figure}[htb]
\centering
\includegraphics*[width=0.4\textwidth]{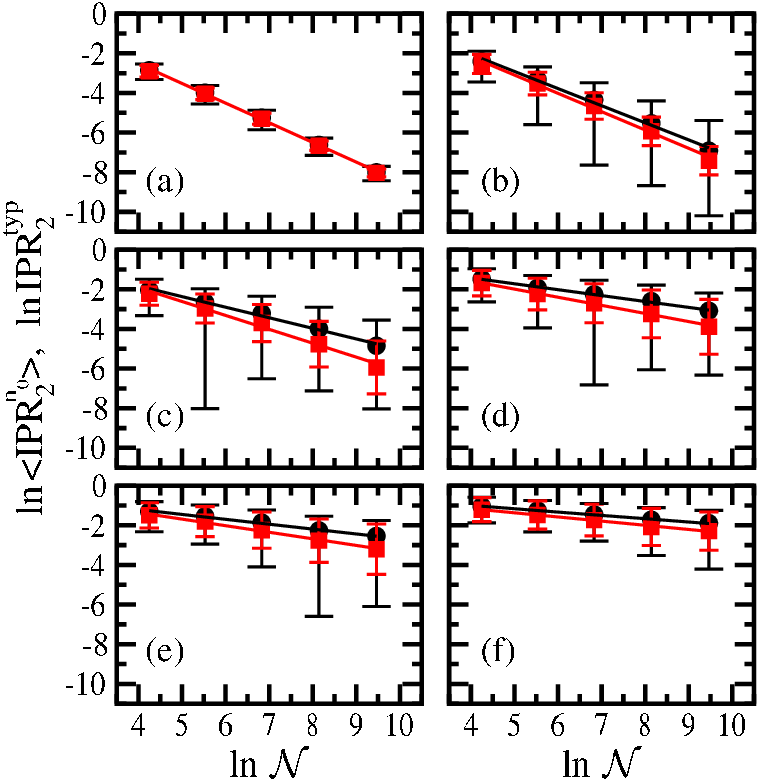}
\caption{(Color online) $ \ln \langle \text{IPR}_2^{n_0} \rangle $ vs  $\ln {\cal N}$ (dark circle) and $ \ln \text{IPR}_2^{\text{typ}}$ vs $\ln {\cal N}$ (light square) for $h=1$ (a), $h=1.5$ (b), $h=2.0$ (c), $h=2.7$ (d), $h=3.2$ (e), and $h=4.0$ (f). Error bars are standard deviations over $10^5$ values of $\text{IPR}_2^{n_0}$ (dark color) or of $\ln \text{IPR}_2^{n_0}$ (light color).}
\label{fig:supp01}
\end{figure}

The dashed lines in Fig.~\ref{fig:SPoneCase} correspond to an algebraic decay described by the generalized dimension, $\langle F(t) \rangle \propto t^{-\widetilde{D}_2}$. As shown in Fig.~\ref{fig:supp01}, $\widetilde{D}_2$ is extracted from the best  linear fit to $\ln \langle \text{IPR}_2^{n_0}  \rangle$ vs $\ln{\cal N}$ for $L=8,10,12,14,16$. 
The error bars are standard deviations over $10^5$ different values of $\text{IPR}_2^{n_0} $ for each $L$.  At small disorder, the error bars are small. As $h$ approaches $h_c$, the dispersion of the values of $\text{IPR}_2^{n_0} $ and therefore the uncertainty in the value of $\widetilde{D}_2$ increases. At very large disorder, the error bars decrease again.

For small disorder, $h\lesssim 1$, the system is still  close to the metallic phase and the decay is diffusive, $\widetilde{D}_2 \sim 1$. In this case, the exponent of the numerical powerlaw decay agrees extremely well with $\widetilde{D}_2$ when the system size is large [see Fig.~\ref{fig:SPoneCase} (a)]. As $h$ increases, $\widetilde{D}_2$ decreases, but not as fast as the numerical exponent. For $h=1.5$ [Fig.~\ref{fig:SPoneCase} (b)], the agreement between the numerical curve and $\langle F(t) \rangle \propto t^{-\widetilde{D}_2}$ is not very good anymore.

In the vicinity of the critical point, Figs.~\ref{fig:SPoneCase} (c) and (d), oscillations are seen approximately in the same time interval of the algebraic decay of Fig.~\ref{fig:SPoneCase} (a). The generalized dimension is now $\widetilde{D}_2 < 1/2$ and it agrees well with the rate of the damping of those oscillations, while the powerlaw decay appears now latter in time.  As $L$ increases, the amplitudes of the oscillations decrease and the slope of the subsequent powerlaw decay becomes more pronounced and closer to $t^{-\widetilde{D}_2}$. It is thus plausible to expect that for very large system sizes, $\widetilde{D}_2$ might be able to capture the algebraic decay also for large disorder. This expectation is further supported by the results below for the  time-averaged survival probability.


\section{Time-Averaged Survival Probability}  
In the analysis of the dynamics of noninteracting systems at the mobility edge~\cite{Ketzmerick1992,Huckestein1994,Huckestein1999,Ng2006}, the commonly employed quantity is the time-averaged survival probability, which smoothes the fluctuations in $\langle F(t) \rangle$. It is defined as, 
\begin{equation}
C(t) \equiv \frac{1}{t} \int_0^t \langle F(\tau) \rangle d\tau  .
\end{equation}
To reduce also the fluctuations in the values of IPR$_2^{n_0}$, one often deals with the so-called typical inverse participation ratio, IPR$_2^\text{typ} \equiv \exp (\langle \ln\text{IPR}_2^{n_0}\rangle)$. The scaling analysis of IPR$_2^\text{typ} $ gives $\widetilde{D}_2^{\text{typ}}$, as shown in Fig.~\ref{fig:supp01}.
The error bars for IPR$_2^\text{typ} $ in that figure are, of course, smaller than those for the regular $\text{IPR}_2^{n_0}$, since now we deal with the dispersions in the values of  $\ln \text{IPR}_2^{n_0} $.

\begin{figure}[htb]
\centering
\includegraphics*[width=3.in]{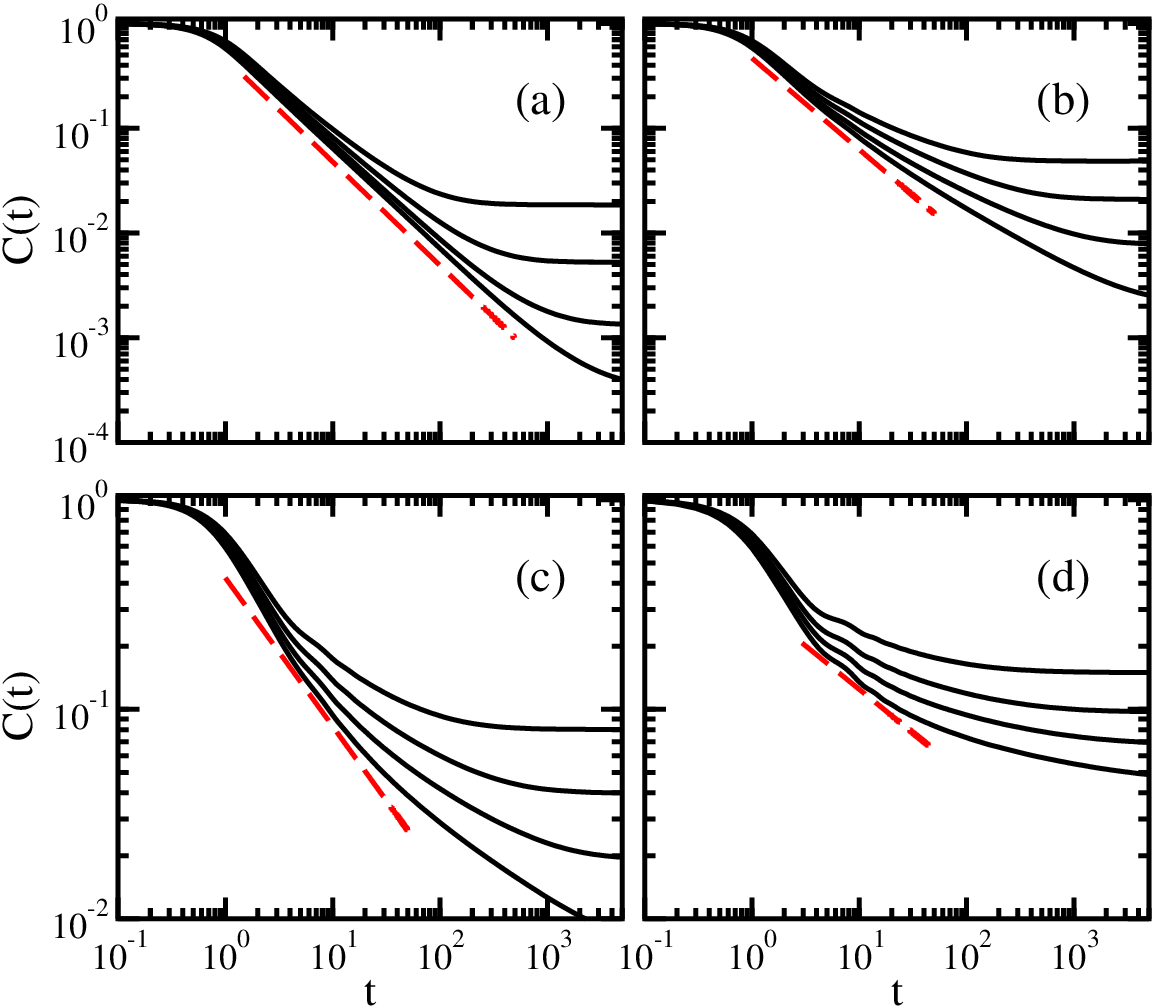}
\caption{(Color online) Time-averaged survival probability for $h=1.0$ (a), $h=1.7$ (b), $h=2.0$ (c), and $h=2.7$ (d) for $L=10, 12, 14, 16$ from top to bottom. Dashed lines gives $t^{-\widetilde{D}_2^{\text{typ}} }$, where $\widetilde{D}_2^{\text{typ}} = 0.99$ (a), $0.87$ (b), $0.72$ (c), and $0.42$ (d). 
}
\label{fig:AverageSP}
\end{figure}

In Fig.~\ref{fig:AverageSP}, we compare $C(t)$ with $t^{-\widetilde{D}_2^{\text{typ}}} $.
When the system is still close to the metallic phase, as in Fig.~\ref{fig:AverageSP} (a), the decay of $C(t)$ is smooth all the way to saturation and in excellent agreement with $t^{-\widetilde{D}_2^{\text{typ}}}$, especially for $L=16$. 

As the disorder increases, the powerlaw exponent decreases, but the short-time dynamics does not change much. This creates an abrupt contrast between the two time scales, resulting in a visible elbow [see Fig.~\ref{fig:AverageSP} (c) and (d)]. 
As $h$ increases, we also notice that the time interval for the agreement between the algebraic decay of $C(t)$ and $t^{-\widetilde{D}_2^{\text{typ}}}$ shortens and starts later in time (compare the four panels). Yet, for a fixed disorder, the agreement also improves with $L$. This indicates that for system sizes larger than available for exact diagonalization, $\widetilde{D}_2^{\text{typ}}$ should be able to describe the powerlaw decay for long times, even when the disorder is strong.

\section{Powerlaw Exponent and System Size} 
The exponent of the powerlaw decay of $F(t)$ contains important information about the system: 

(i) Because it coincides with the generalized dimension $\widetilde{D}_2$, it indicates the level of delocalization of the initial state. Since $\widetilde{D}_2 \sim D_2$, as  suggested in \cite{Huckestein1997} and confirmed by us for our model, the powerlaw exponent also manifests the level of delocalization of the eigenstates.

(ii) It gives information about the correlations between the components $|C^{\alpha}_{n_0} |^2$, because the algebraic decay implies that  \cite{Huckestein1994,Chalker1988,Kravtsov2011},
\begin{eqnarray}
&&\langle F(t) \rangle = \left\langle \sum_{\alpha,\beta} |C^{\beta}_{n_0} |^2 |C^{\alpha}_{n_0} |^2 e^{i (E_{\beta} - E_{\alpha} ) t} \right\rangle \nonumber \\
&& \xrightarrow{\omega = E_{\beta} - E_{\alpha}} \int_{-\infty}^{\infty} d \omega e^{i \omega t} |\omega|^{\widetilde{D}_2 - 1} \propto t^{-\widetilde{D}_2} .
\label{eq:correlations}
\end{eqnarray}

When the eigenstates, and consequently the initial state, are extended and thus similar to random vectors, as it happens in the chaotic domain ($h\lesssim 1$) for states close to the middle of the spectrum, the components $|C^{\alpha}_{n_0} |^2$ are uncorrelated random numbers. In this case $\text{IPR}_2^{n_0} \propto {\cal N}$ and the dynamics is  diffusive ($\widetilde{D}_2 \sim 1$), as obtained also in  \cite{Luitz2014}. As the disorder increases, the states become multifractal; the components $|C^{\alpha}_{n_0} |^2$ show large fluctuations and become gradually more correlated, so $\text{IPR}_2^{n_0} \propto {\cal N}^{\widetilde{D}_2}$ with $\widetilde{D}_2 < 1$, resulting in a subdiffusive dynamics. The limited spreading of the initial state quantified by $\widetilde{D}_2$ reflects, as made explicit by Eq.~(\ref{eq:correlations}),  the level of correlations between the components  $|C^{\alpha}_{n_0} |^2$.

\begin{figure}[htb]
\centering
\includegraphics*[width=3.in]{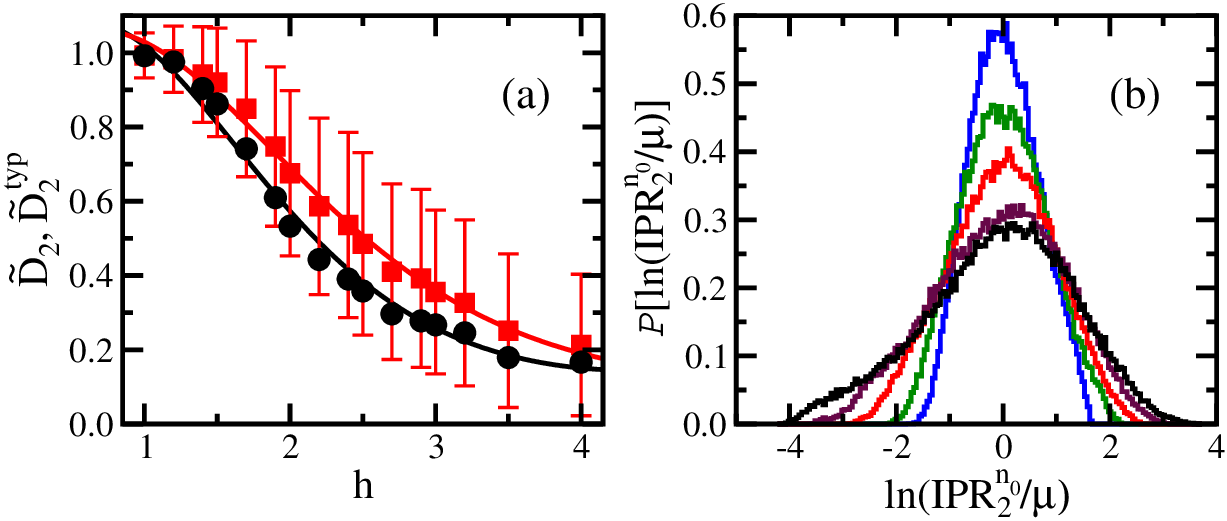}
\caption{(Color online) $\widetilde{D}_2 $ (dark circle) and $\widetilde{D}_2^{\text{typ}} $ (light square) vs disorder strength (a)  and the distribution of $\ln(\text{IPR}_2^{n_0}/\mu)$ for $h=2.7$ and $L=8, 10, 12, 14, 16$ from highest to lowest peak (b). In (a): solid lines are fitting curves. Only the error bars for $\widetilde{D}_2^{\text{typ}} $ are shown. They are smaller than those for $\widetilde{D}_2 $, in accordance with the standard deviations in Fig.~\ref{fig:supp01}.
}
\label{fig:Distribution}
\end{figure}

Figure~\ref{fig:Distribution} (a) shows how $\widetilde{D}_2 $ and $\widetilde{D}_2^{\text{typ}} $ depend on the disorder strength. In parallel with the standard deviations in Fig.~\ref{fig:supp01}, the error bars are larger for $\widetilde{D}_2 $ than for $\widetilde{D}_2^{\text{typ}} $. We show only the latter to simplify the figure. Within errors, the two generalized dimensions coincide. As $h$ increases,  the number of states that contribute to the evolution of the initial state shrinks and the generalized dimensions decrease. The decay is evident for $1<h<4$ and it becomes extremely slow afterwards. We avoid an analysis of what happens for $h>4$, because for the very small system sizes $L=8,10$ we actually see an approach to on-site localization. 

Even though it is not clear at this point how to identify the MBL critical point from Fig.~\ref{fig:Distribution} (a), a comparison with previous studies is instructive. The values of $h$ for the mid-point between a metal and an insulator, that is $\widetilde{D}_2^{\text{typ}} \sim 1/2$ ($h\sim 2.5$), and for the inflection point of the fitting curve ($h\sim 2$) are not too far from the critical points found in \cite{Pal2010,DeLuca2013}. In addition, the point  of an almost halt in the decay of the values of the generalized dimensions, $h\sim 4$, is very close to the critical point $h_c\sim 3.7$ obtained in Ref.~\cite{Luitz2014} for states that, as in our case, live close to the middle of the spectrum.

We expect $\widetilde{D}_2 $ and $\widetilde{D}_2^{\text{typ}} $ to get closer for scaling analysis performed with larger system sizes than the very few ones now available for exact diagonalization. As $L$ increases, they should also better agree with the powerlaw exponent of $F(t)$. These claims find support already in the results for $L=14$ and $16$.
If the powerlaw behavior is indeed to be described by the generalized dimension, then the algebraic decays for different system sizes must coincide. In Figs.~\ref{fig:SPoneCase} and \ref{fig:AverageSP}, the slopes are visibly different for small $L$'s, but they get closer for $L=14$ and $16$. This suggests that the scaling analysis should become more accurate for sizes $L>14$.

Figure~\ref{fig:Distribution} (b) endorses the proximity of the results for $L=14$ and $16$. It shows the distribution of the inverse participation ratios. IPR$_2^{n_0}$ fluctuates with disorder realization and initial state. However, the validity of  Eq.~(\ref{eq:D2forInitial}) 
presupposes that $\widetilde{D}_2 $ does not depend strongly on what is used on the left side of that equation, whether it is  $\langle \text{IPR}_2^{n_0}\rangle$, IPR$_2^\text{typ}$, or the most probable value of IPR$_2^{n_0}$. This implies that the distribution of IPR$_2^{n_0}$ normalized to its median $\mu$ must have a scale invariant shape~\cite{Evers2000,Evers2008}.
As seen in Fig.~\ref{fig:Distribution} (b), the distribution $P[\ln (\text{IPR}_2^{n_0}/\mu)]$  broadens considerably from $L=8$ to $12$, but the shapes are similar for $L=14$ and $16$. 

In noninteracting disordered systems described by the powerlaw random banded matrix, numerical evidence for the scale invariance of $P[\ln \text{IPR}_2^{n_0}]$ was achieved~\cite{Evers2000} already for ${\cal N} \gtrsim 300$, in contrast with the ${\cal N} \gtrsim 3000$ needed here. The existence of more correlations between the matrix elements of our system when compared to random matrices may justify such large difference. The number of nonrandom elements in the Hamiltonian matrix of  Eq.~(\ref{ham}) is much larger than in noninteracting systems, such as those described by the tight-binding model or  the powerlaw random banded matrix. 

\section{Conclusion}  
We studied the dynamics of an isolated disordered 1D Heisenberg model as it approaches the MBL phase. The analysis was based on the entire evolution of the survival probability $F(t)$, from $t=0$ to $t \rightarrow \infty$, for initial states corresponding to site-basis vectors. $F(t)$ is one of the simplest quantities that can reveal the multifractality of the eigenstates. It also appears explicitly in the evolution of observables~\cite{TorresALL}.

The dynamics of clean and disordered interacting systems is comparable at short times. For both, the Gaussian decay rate of $F(t)$ coincides with the width of the LDOS. In the presence of disorder, the LDOS gets sparse, reflecting the reduced number of states participating in the dynamics and the multifractality of the eigenstates. As a result, the behavior of $F(t)$ at long times becomes powerlaw. 

The exponent of the powerlaw decay coincides with the generalized dimension $D_2$. This finding establishes a parallel with previous works about the dynamics of noninteracting systems at criticality and may help advance our understanding of transport properties in interacting systems. It also implies that from $F(t)$, one can infer the level of delocalization of the initial states and eigenstates, as well as the correlations of their components. This is advantageous, since numerical methods other than exact diagonalization are available for studying dynamics, which gives access to larger system sizes. The dynamics can also be studied experimentally with quantum simulators. 


\begin{acknowledgments}
This work was supported by the  NSF grant No.~DMR-1147430. EJTH acknowledges support from CONACyT, Mexico. LFS thanks the ITAMP hospitality, where part of this work was done. We thank J.~A.~Mendez-Bermudez and Z.~Papi\'{c} for useful suggestions.
\end{acknowledgments}



\begin{thebibliography}{61}
\bibitem[{not({\natexlab{b}})}]{notePR}
\bibinfo{note}{The quantity that we call here participation ratio is also
  referred to in the literature as inverse participation ratio. We save the
  latter for 1/P$_2$.}
\end{thebibliography}
\end{document}